\documentclass[useAMS,usegraphicx,usenatbib,galley]{mn2e}
\usepackage{graphicx}

\def\mearth{\ifmmode {\rm M_{\oplus}}\else $\rm M_{\oplus}$\fi}
\def\Mearth{\ifmmode {\rm M_{\oplus}}\else $\rm M_{\oplus}$\fi}
\def\Rearth{\ifmmode {\rm R_{\oplus}}\else $\rm R_{\oplus}$\fi}
\def\Ms{\ifmmode {M_s}\else $M_s$\fi}
\def\Mp{\ifmmode {M_p}\else $M_p$\fi}
\def\Rp{\ifmmode {R_p}\else $R_p$\fi}
\def\rearth{\ifmmode {\rm R_{\oplus}}\else $\rm R_{\oplus}$\fi}

\newcommand{\Mdot}{\dot M_*}
\newcommand{\Mdotzero}{\dot M_{*0}}

\newcommand{\Mdotnorm}{\dot M_{\rm *,norm}}
\newcommand{\Mstar}{M_*}
\newcommand{\Mdisk}{M_{\rm d}}
\newcommand{\Mdiskthin}{M_{\rm d,thin}}
\newcommand{\Mdiskfit}{M_{\rm d,fit}}
\newcommand{\Msun}{M_{\odot}}

\newcommand{\Rstar}{R_*}
\newcommand{\Rsun}{R_\odot}
\newcommand{\Msunperyr}{M_{\odot}\,{\rm yr}^{-1}}
\newcommand{\Lsun}{L_{\odot}}
\newcommand{\Lstar}{L_*}

\title[Demographics of Oph and Tau Transition Discs]
{Demographics of Transition Discs in Ophiuchus and Taurus}
\author[J.\ R.\ Najita, S.\ M.\ Andrews, and J.\ Muzerolle]{Joan R. Najita$^{1,2}$\thanks{E-mail:
najita@noao.edu (JRN)}, Sean M.\ Andrews$^{2}$, and James Muzerolle$^{3}$ \\
$^{1}$National Optical Astronomy Observatory, 950 Cherry Avenue, Tucson, AZ. 85719, USA\\
$^{2}$Harvard-Smithsonian Center for Astrophysics, 60 Garden Street, Cambridge, MA 02138, USA \\
$^{3}$Space Telescope Science Institute, 3700 San Martin Drive, Baltimore, Maryland 21218, USA}
\begin{document}

\date{}

\pagerange{\pageref{firstpage}--\pageref{lastpage}} \pubyear{2015}

\maketitle

\label{firstpage}

\begin{abstract}
Transition disc systems are young stars 
that appear to be on the verge of dispersing their protoplanetary discs. 
We explore the nature of these systems by 
comparing the stellar accretion rates $\Mdot$ and disc masses $\Mdisk$
of transition discs and normal T Tauri stars in Taurus and Ophiuchus. 
After controlling for the known dependencies of $\Mdot$ and
$\Mdisk$ on age, 
$\Mdot$ on stellar mass, and $\Mdisk$ on the presence of
stellar or sub-stellar companions,
we find that the normal T Tauri stars show a trend of $\Mdot$
increasing with $\Mdisk$.
The transition discs tend to have higher average disc masses than
normal T Tauri stars as well as lower accretion rates
than normal T Tauri stars of the same disc mass.
These results are most consistent with the interpretation that the
transition discs have formed objects
massive enough to alter the accretion flow, i.e., single or multiple
giant planets.
Several Ophiuchus T Tauri stars that are not known transition disc
systems  
also
have very low accretion rates for their disc masses.
We speculate on the possible nature of these sources.
\end{abstract}


\section{Introduction}
\label{sec: intro}

The population of young stellar objects known as transition discs
have unusual spectral energy distributions (SEDs)
that suggest that some portion of the inner disc has become
optically thin
(Strom et al.\ 1989; Skrutskie et al.\ 1990).
The radial distribution of the dust disc has been inferred from
detailed modeling of the SEDs, which
have significant excesses beyond $10\micron$, like those of normal T Tauri stars,
but weak excesses at mid-infared wavelengths
(e.g., D'Alessio et al.\ 2005; Calvet et al.\ 2005).
The optically thin region inferred from the modeling---a
central ``hole'' in the dust distribution or an annular ``gap''
within a radius $r_h$ of 15\,AU to $>50$\,AU depending on the source---has
been confirmed by submillimeter interferometric imaging of
bright, nearby systems (Andrews et al.\ 2011).
These properties suggest that transition discs are in the process of
dispersing their discs (see Espaillat et al.\ 2014 for a review).

Several scenarios have been proposed to explain the properties of
transition discs.
An early suggestion was that grain growth and planetesimal formation,
the first steps toward planet formation via core accretion,
might reduce the optical depth
of the inner disc by reducing the dust cross-sectional area
(Strom et al.\ 1989; see also Dullemond \& Dominik 2005; Tanaka et al.\ 2005).
The formation of giant planets is also
expected to carve a gap or inner cavity in the disc,
rendering some or all of the inner disc optically thin
(Skrutskie et al.\ 1990; Marsh \& Mahoney 1992, Lubow et al.\ 1999; 
Calvet et al.\ 2002;
Rice et al.\ 2003; Quillen et al.\ 2004, Calvet et al.\ 2005,
D'Alessio et al.\ 2005;
Rice et al.\ 2006).
Alternatively,
a photoevaporative wind, launched by the irradiation of the disc surface by
stellar UV and X-rays, may also deplete the inner disc,
by limiting the ability of material accreting from the outer disc to
resupply the material in the inner disc
(Clarke et al.\ 2001; Alexander et al. 2006;
Alexander \& Armitage 2007; Owen et al.\ 2010, 2011; Gorti et
al.\ 2009).
These interpretations are not mutually exclusive.  For example,
photoevaporation
may disperse a disc more rapidly once a gap forms as a result
of giant planet formation (e.g., Rosotti et al.\ 2013).
Transition discs may also be produced through multiple pathways,
with different processes potentially dominating on different
timescales (e.g., Williams \& Cieza 2011).

Because multiple processes can produce transition disc SEDs,
other information (beyond SEDs) has been used in efforts to distinguish
among the potential explanations. Detailed studies of individual
transition discs have reported evidence for an orbiting companion
(e.g., Kraus \& Ireland 2012; Huelamo et al.\ 2011)
or its circumplanetary disc
(Brittain et al.\ 2013, 2014).
Other studies have probed the spatial and velocity distribution
of disc gas to look for evidence of photoevaporative outflows
(e.g., Pascucci \& Sterzik 2009; Sacco et al.\ 2012)
or the radial truncation of the gaseous disc
(e.g., Najita et al.\ 2008; Dutrey et al.\ 2008). The latter
is not expected in the planetesimal scenario.

An alternative way to
distinguish among the possible scenarios
is to compare the demographic properties
of populations of transition discs with those of
normal (non-transition) T Tauri stars.
Whereas
recent studies have employed stellar accretion rates
alone or in combination with properties such as X-ray luminosity and stellar
mass for this purpose (Kim et al.\ 2009, 2013; Espaillat et al.\ 2012; Fang et al.\ 2013;
Manara et al.\ 2014),
the combination of stellar accretion rates and disc masses
has also been suggested as a potentially powerful way to distinguish
among the scenarios described above
(Najita et al.\ 2007; Alexander \& Armitage 2007;
see also Cieza et al.\ 2008, 2010, 2012; Mendigut\'ia et al.\ 2012).
To summarize the basic idea:

\noindent (a) In the grain and planetesimal growth scenario,
the dust component of the disc within $r_h$
is altered but not the gas; the stellar accretion rate
is therefore expected to be similar to that of normal T Tauri stars.
Because planet formation is a common
outcome of disc evolution (e.g., Fressin et al.\ 2013; Dong \& Zhu
2013; Petigura et al.\ 2013), and planetesimal formation is the
first step toward planet formation via core accretion, planetesimals
are expected to form in discs of all masses.

\noindent (b) In contrast, when a forming Jovian-mass planet clears a gap in
the disc, most of the accretion flow from the outer disc
accretes onto the planet (in streams), and the stellar accretion rate is
reduced to $\sim 0.1$ of its original T Tauri value
(Lubow et al.\ 1999; but see also Zhu et al.\ 2011).
Because a Jovian-mass planet
represents a significant mass reservoir relative to the median T Tauri
disc mass (Andrews et al.\ 2013), planets massive enough to open a
gap (e.g., $\ga 1 M_J$) are expected to form in discs with higher
than average masses.

\noindent (c) The accreting planet could
eventually reach a mass high enough ($\sim 5 M_J$)
to suppress flow past the planet.  When the remaining inner
disc material accretes on to the star, an inner cavity
in the gas and dust distribution of the disc would remain, with no
further stellar accretion (e.g., Lubow et al.\ 1999; Lubow \& D'Angelo 2006).
Forming such high mass planets is favored in higher mass discs.

\noindent (d) If a photoevaporative wind can remove material from the disc faster
than it can be replenished by accretion, the inner disc is starved of
gas and dust, and stellar accretion drops to a low or negligible rate.
This situation is more readily achieved in low mass discs
(all other factors being equal; e.g., stellar irradiation fields)
with initially modest accretion rates (e.g., Clarke et al.\ 2001;
Alexander \& Armitage 2007; Owen et al.\ 2010).

A previous study of stellar accretion rates and disc masses
found evidence for a modest trend between stellar accretion rate
and disc mass among single, normal T Tauri stars in Taurus
(Najita et al.\ 2007).
In addition, the Taurus transition discs were found to have
disc masses $\sim 4$ times larger than the normal T Tauri stars
and stellar accretion rates $\sim 10$ times
lower than those of normal T Tauri stars of the same disc mass.
These properties are most consistent
with the interpretation that the Taurus transition discs are forming
Jovian mass planets.

Here we extend the Taurus study by examining the
stellar accretion rates $\Mdot$ and disc masses $\Mdisk$ of
transition discs and T Tauri stars in Ophiuchus
in combination with the earlier Taurus results.
Ophiuchus is one of the few star forming regions in which
disc masses are available for
a significant number of young stellar objects.
Our study differs in several ways from the approach 
taken by Najita et al.\ (2007). The transition disc sample 
studied here is restricted to sources with evidence for large 
inner holes, i.e., sources with either 
a large MIR dip in the SED or 
a resolved cavity in submillimeter imaging of the source. 
Because of the large range of stellar spectral types in the 
Ophiuchus sample, we also correct for the 
dependence of stellar accretion rate on stellar mass. 
Our data sets are described in \S2 and our analysis and results in \S3.
In \S4, we discuss the location of transition discs in the
$\Mdot$--$\Mdisk$ plane and compare our results to those of previous
studies. Our conclusions are summarized in \S5.

\section{Disk Masses and Stellar Accretion Rates}
\label{sec: data}

For both the Taurus (Tau) and Ophiuchus (Oph) samples,
we use stellar accretion rates and disc masses from
the literature, with a few emendations as described below.
Because we intend to compare the results for Oph
with those previously obtained for Tau,
the disc masses and stellar accretion rates for the
two samples are treated as similarly as possible.

Millimeter continuum fluxes have been
reported previously for many Oph sources.
To incorporate these measurements in a homogeneous way in
deriving disc masses,
we follow the approach of Andrews et al.\ (2013)
in their analysis of Tau disc masses.
For each disc, we compiled and fit, using a power law model,
all the available continuum measurements in the literature
at wavelengths from $700\micron$ to 3\,mm
(Andre \& Montmerle 1994; Dent et al.\ 1998; Nuernberger et al.\ 1998;
Motte et al.\ 1998; Stanke et al.\ 2006; Andrews \& Williams 2007a, b;
Patience et al.\ 2008; Andrews et al.\ 2009, 2010; Ricci et al.\ 2010).
These model fits and their associated uncertainties were
extrapolated to estimate a flux density at a fiducial wavelength
of 1.33\,mm.  If there was only 1 datapoint available
in this range,
we applied a mean power-law scaling to estimate the 1.33\,mm flux density,
where $F_{\nu} \propto \nu^{2.5\pm0.4}$.

We then calculated disc masses assuming the standard optically thin,
isothermal approach, i.e.,
\begin{equation}
\Mdiskthin = {{d^2 F_\nu}\over {\kappa_\nu  B_\nu(T_d)}},
\end{equation}
for a fixed distance $d$ (125\,pc for Oph, 140\,pc for Tau) and
opacity ($\kappa_{1.33mm (dust)} = 2.3\,{\rm cm^2/g}$;
dust-to-gas ratio = 0.01).
In these calculations, the average dust temperature for
each source was computed from a simple estimate of the
incident irradiation field, such that
$T_d  = 25\,{\rm K} (\Lstar/\Lsun)^{1/4}.$
This approximation assumes that the dust in discs is located at a
common effective orbital distance, which is a
reasonable assumption when spatially
resolved information is unavailable
(see Andrews et al.\ 2013 for details).

For Tau, we adopted the disc masses tabulated by Andrews et al.\ (2013)
using the above method.
For Oph, we determined $\Mdiskthin$
using stellar luminosities from McClure et al.\ (2010).
For both the Tau and Oph samples, we also investigated the impact of adopting $\Mdisk$
values estimated from fitting the SED and resolved millimeter imaging data $\Mdiskfit$.
These are available for the brighter Tau and Oph sources
(Andrews \& Williams 2007b; Andrews et al.\ 2009, 2010, 2011).

The primary source of Oph stellar accretion rates
is the large sample reported by Natta et al.\ (2006).
Because of the high extinction toward $\rho$\,Oph, Natta et al.\
determined accretion rates from the IR hydrogen lines (Pa\,$\beta$ or
Br\,$\gamma$) using the calibration of Muzerolle et al.\ (1998)
to convert the IR line luminosities to accretion luminosities.
For consistency with the adopted Oph disc masses,
which involve a correction for $L_*$,
we rederived the stellar accretion rate for each Oph source
from the Pa\,$\beta$ equivalent width and $J$-band magnitude
reported by Natta et al.\ (2006) using the
extinction and $L_*$ from McClure et al.\ (2010).
The stellar masses used in the accretion rate determination
were derived from the Siess et al.\ (2000)
stellar evolutionary models using the approach and
assumptions adopted by Andrews et al.\ (2013).

A few additional Oph accretion rates were adopted from
Espaillat et al.\ (2010),
Eisner et al.\ (2005), and
Valenti et al.\ (1993).
The accretion rate determined by Espaillat et al.\ (2010) for
DoAr44 (also known as ROXs-44)
was derived from the $U$-band excess reported by Bouvier \& Appenzeller (1992)
following the Gullbring et al.\ (1998) method (Andrews et al.\ 2011;
C. Espaillat 2013, personal communication).
Eisner et al.\ (2005) report an accretion rate for WaOph6 (V2508 Oph)
that is based on a $U$-band excess derived from $UBVRI$ photometry
and high resolution optical spectroscopy following the method of
Gullbring et al.\ (1998).
Because the Muzerolle et al.\ (1998) relations used by
Natta et al.\ (2006) were originally calibrated with accretion rates
derived via the Gullbring method, the set of $\Mdot$ values used
here should be on a consistent footing.
The properties of the Oph sample are shown in Table 1.

For the Tau sample, we adopted stellar accretion rates
from Najita et al.\ (2007)
and also included
a few additional stellar accretion rates from
Valenti et al.\ (1993; GI Tau, HN Tau) and
Hartigan et al.\ (1995; HK Tau).
As in Najita et al.\ (2007), the values from Hartigan et al.\ (1995)
were scaled downward
by a factor of 0.17 for consistency with the stellar accretion rates of
Hartmann et al.\ (1998).\footnote{In Najita et al.\ (2007) we also scaled
the stellar accretion rates of White \& Ghez (2001) up by 2.4.
}
In order to investigate the impact of the known trend between
$\Mdot$ and $\Mstar$,
we also adopted the stellar masses reported by
Andrews et al.\ (2013) using the Siess et al.\ (2000) models.
The properties of the single star Tau sample are shown in Table 2.

The Oph sample includes both sources in the $\sim 2$\,Myr old surface
population and the surrounding region (Wilking et al.\ 2005)
as well as the younger embedded sources in the Oph core ($< 1$\,Myr;
Luhman \& Rieke 1999).
In comparison, Tau has a median age of $\sim 2$\,Myr with a
distribution between 1\,Myr and a few Myr (Kenyon \& Hartmann 1995;
Hartmann 2001; Luhman et al.\ 2003; Andrews et al.\ 2013).

\section{Results}
\label{sec: results}

Despite the large samples of Oph stellar accretion rates
and disc mass estimates reported by
Natta et al.\ (2006) and Andrews \& Williams (2007a) ($\sim$140 sources
in each) the overlap between the two samples is limited
($\sim 40$ sources).
The limited overlap is partly due to different areal coverage.
Natta et al.\ (2006) studied the Oph core, whereas Andrews \& Williams (2007a)
studied a larger area. In addition, both samples include a significant
fraction of sources that are either Class III or Class I (25\%--35\%).
Figure 1 shows sources that have
detections or upper limits (indicated by arrows)
for both quantities.
Sources with a known stellar
companion within $1\arcsec$ (Reipurth \& Zinnecker 1993;
Barsony et al.\ 2005; Ratzka et al.\ 2005)
are shown as open circles.
Sources without such a companion are regarded as single stars
(solid symbols).
The binaries have lower average disc masses than the
single stars, consistent with similar trends noted elsewhere
in the literature
(e.g., Jensen et al.\ 1994, 1996; Andrews \& Williams 2005;
Harris et al.\ 2012).
The presence of a stellar companion is believed to
dynamically alter the structure of the disc, reducing its mass.
We therefore excluded the binaries from further consideration.

In Figure 1, sources with later spectral types (later than
M1) have lower average $\Mdot$ than earlier types,
a result that
is not surprising given the well-known trend of increasing
$\Mdot$ with $\Mstar$
(e.g., Muzerolle et al.\ 2003; Calvet et al.\ 2004; Natta et al.\ 2006;
Fang et al.\ 2009; Rigliaco et al.\ 2011; Manara et al.\ 2012).
This trend was not an issue in the earlier Taurus study,
because the Tau sample spans a more
limited range of spectral types
(K2--M2; $\sim 80$\% are K3-M0 or
$\sim 0.5-1.5\Msun$).
Because the trend of $\Mdot$ with $\Mstar$ has a large scatter
at $\sim 1$\,Myr age ($\sim 2$ orders of magnitude at a given
$\Mstar$; e.g., Fang et al.\ 2013), the trend is only apparent
over a mass range of $\ga 1$\,dex (e.g., Muzerolle et al.\ 2003;
Manara et al.\ 2012), as in our Oph sample
(A1--M6; $\sim 0.07-4\Msun$).

\setcounter{figure}{0}
\begin{figure}
\includegraphics[width=84mm]{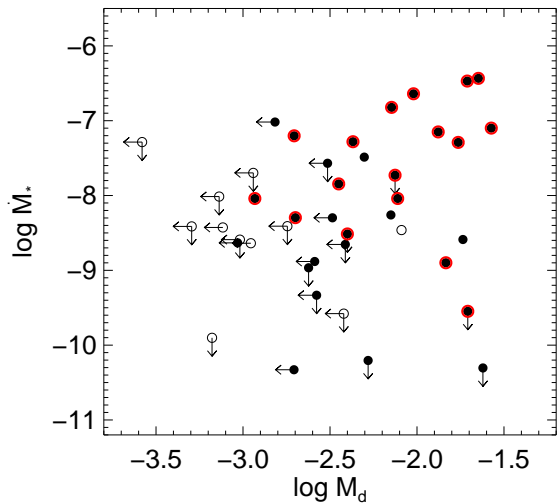}
\caption{Stellar accretion rates (in $\Msunperyr$) and 
disc masses (in $\Msun$) for Ophiuchus sources.
Arrows indicate upper limits.
Closed (open) circles designate sources without (with) known
companions within $1\arcsec$.
Binaries have lower disc masses than single stars.
Sources with spectral types earlier than M0 
(symbols with a red rim) have higher average stellar accretion 
rates than sources with later spectral types. 
}
\end{figure}

We therefore corrected the Oph sample for the trend between
$\Mdot$ and $\Mstar$ in order to include its larger range of
spectral types and maximize its sample size.
The derived Oph accretion rates $\Mdotzero$ were rescaled as
$\log \Mdotnorm = \log \Mdotzero  - 1.3 \log(\Mstar/0.7\Msun)$
using the slope from Manara et al.\ (2012) for a 1\,Myr population
and our estimated stellar masses (section 2).
In exploring alternative approaches, we found that 
the effect of the rescaling on our results is insensitive to
the details of which stellar evolutionary tracks are used in
estimating stellar masses or whether the
appropriate relation has a steeper slope
(e.g., $\Mdot \propto \Mstar^2$ as in Muzerolle et al.\ 2003).
We also explored but did not adopt a correction for a 
possible trend of $\Mdisk \propto \Mstar$ (Andrews et al.\ 2013), 
because it did not affect our results. 
This is understandable because $\Mdisk$ and $\Mstar$ are not 
strongly correlated over the primary mass range 
of our sample ($\log\Mstar$= -0.65 to 0.3; Andrews et al\ 2013).  

Figure 2a shows the rescaled Oph accretion rates plotted against
$\Mdiskthin$ (black dots)
where, for clarity, sources with upper limits on their disc mass
are not shown.
Sources with a transition disc SED and/or a submillimeter cavity
(Espaillat et al.\ 2010, McClure et al.\ 2010; Andrews et al.\ 2011)
are indicated by blue `+' symbols.
Single T Tauri stars (gray dots) and transition discs (blue `x' symbols)
in Taurus are also shown (see Najita et al.\ 2007 and section 2),
where we have applied the same rescaling of $\Mdot$ as a function
of $\Mstar$ that was used for the Oph sample.
The Oph transition discs fall between the regions occupied by the
Tau transition discs and the normal T Tauri stars.

Most of the sources marked as transition discs in Table 1
and Figure 2 have prominent mid-infrared (MIR) dips in their SEDs.
Because there are only 2 such Oph objects with $\Mdot$
measurements
(SR21A, DoAr44; McClure et al.\ 2010),
we also included Oph sources that have
a submillimeter cavity (but lack a MIR SED dip)
for which $\Mdot$ is available
(SR24S, WSB60; Andrews et al.\ 2011).
These sources
(both with and without MIR SED dips) have submillimeter cavity
radii of 15--36\,AU (Andrews et al.\ 2011).
Because submillimeter imaging studies 
have been limited to the brighter sources 
($> 75$\,mJy at $880\micron$ or $\Mdisk \ga 0.005\Msun$ 
for the Oph SMA sample; Andrews et al.\ 2010), the sample of 
cavity sources may be artificially biased to higher disk masses; 
i.e., lower mass disks may possess cavities that have not yet 
been identified.
In contrast, 
selecting transition discs by their prominent MIR SED dip 
should not introduce a similar bias. 
For example, a system like CoKu Tau/4, which has a 
prominent MIR SED dip, has a very 
low disk mass ($\sim 0.001\,\Msun$; D'Alessio et al.\ 2005).

The Tau transition disc sample differs from that of
Najita et al.\ (2007), who defined transition discs more
broadly based on weak or no excess shortward
of $10\micron$ and a significant excess at longer
wavelengths. Because such a broad definition
is debated, we have adopted a more conservative definition here
by selecting only objects with prominent MIR dips in their SEDs
(DM Tau, GM Aur, UX Tau, LkCa15; Calvet et al.\ 2005;
Espaillat et al.\ 2010).
The remaining sources in the Tau sample are classified as
normal T Tauri stars.

\setcounter{figure}{1}
\begin{figure}
\includegraphics[width=82mm]{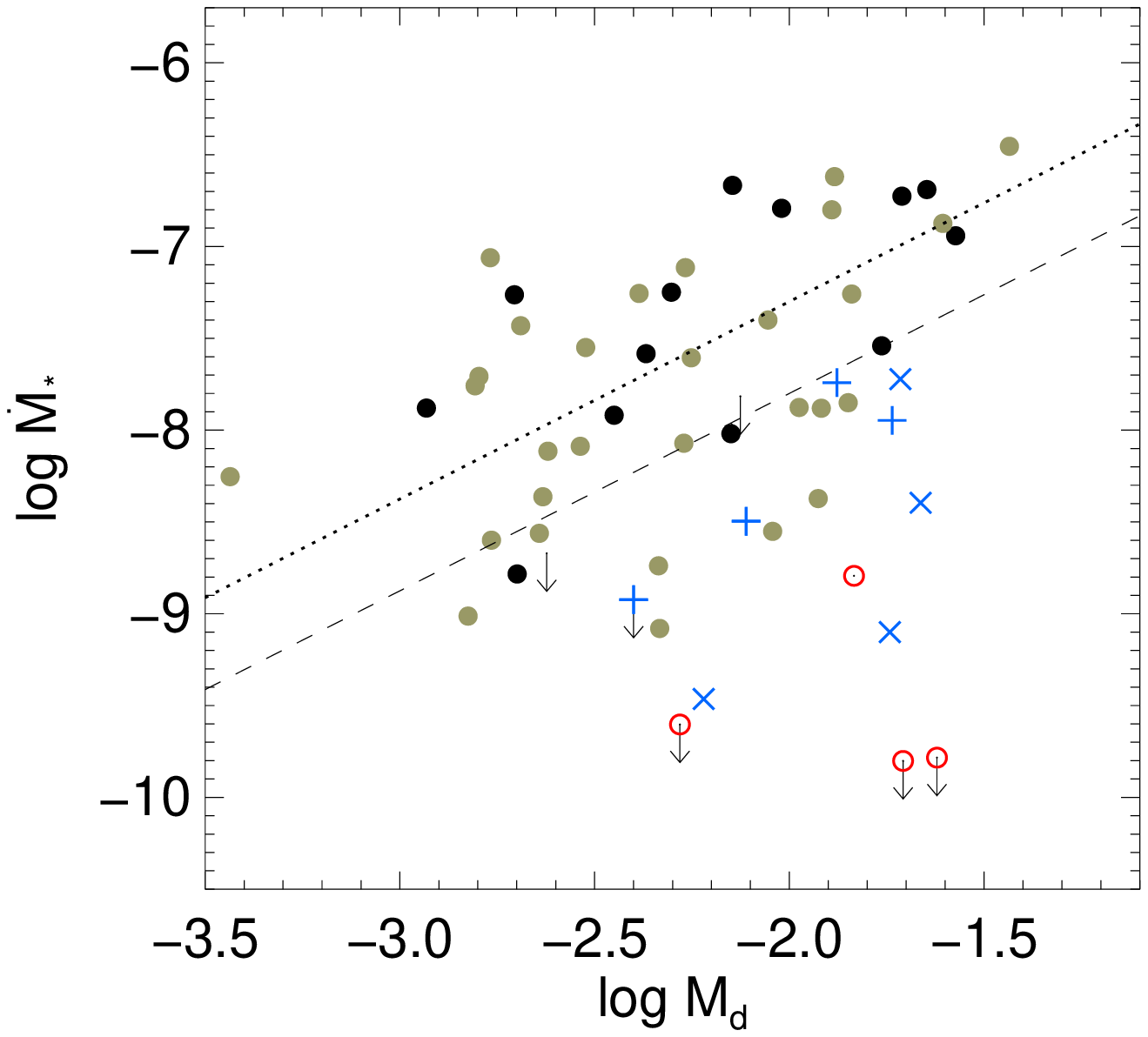}
\end{figure}

\setcounter{figure}{1}
\begin{figure}
\includegraphics[width=82mm]{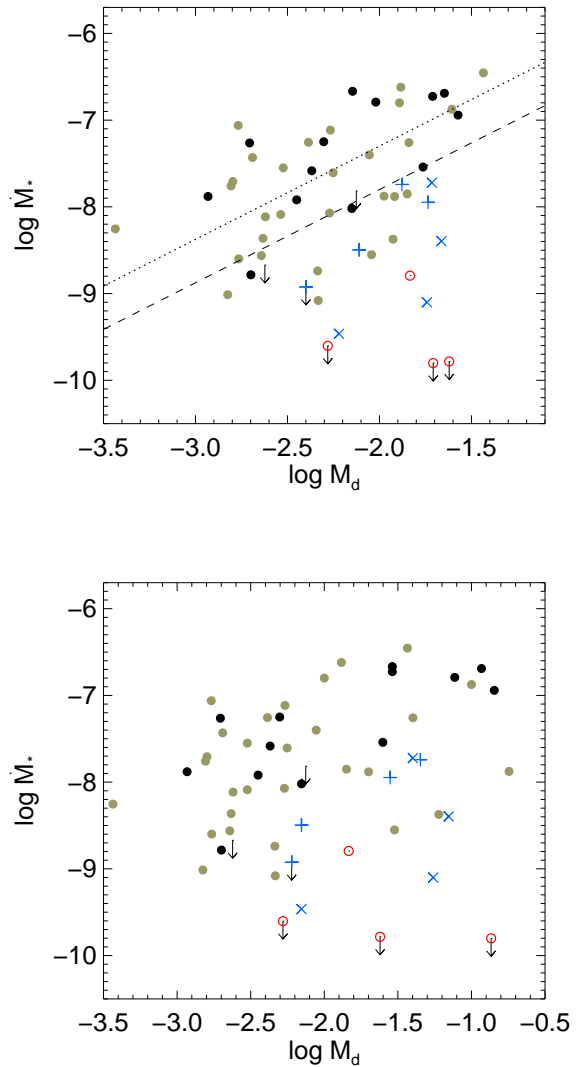}
\caption{Top (a):
Stellar accretion rates (in $\Msunperyr$) and disc masses (in $\Msun$) of
T Tauri stars in Ophiuchus (black dots) and Taurus (gray dots) compared
with transition discs in Ophiuchus (blue '+') and Taurus (blue 'x').
Red circles indicate the Ophiuchus outlier population, i.e.,
T Tauri stars with normal SEDs that have very low accretion rates
for their disc masses.
For clarity, only non-binaries and sources with $\Mdisk$ detections are shown.
Stellar accretion rates have been adjusted for the known dependence of
$\Mdot$ on $\Mstar$.
Arrows indicate sources with upper limits on their stellar accretion rates.
The dotted line is a linear fit to the normal T Tauri stars;
the dashed line, located 1-$\sigma$ below the dotted line, divides the
plane into two regions (see text for details).
Bottom (b):
Similar to the top panel except that disc mass estimates from
fits to resolved millimeter-wavelength visibilities $\Mdiskfit$
are used when available.
In either plot, transition discs have lower than average
accretion rates for their disc mass.
}
\end{figure}

To explore the impact of our simple prescription for disc
mass (eq.~1), Figure 2b shows the same information except
that the Tau and Oph disc masses estimated
from radiative transfer models that simultaneously fit SEDs and
resolved millimeter-wavelength visibilities $\Mdiskfit$ are used
when available
(Andrews \& Williams 2007b; Andrews et al.\ 2009, 2010, 2011).
Of the 13 Oph sources with $\Mdiskfit$ values,
roughly half have $\Mdiskfit$ values within 60\% of the
$\Mdiskthin$ value.  The other half
have $\Mdiskfit$ values $\sim 5$ times larger.
The large difference for sources such as WaOph6 and DoAr25
occurs because the dust associated with these sources is
colder and the disc is larger in size
than one would infer from the simple optically thin scaling
used for $\Mdiskthin$.
A similar result is found for the Tau sources.

Note that the Tau accretion rates and disc masses shown
in Figure 2a differ from those adopted in
Najita et al.\ (2007). Here the accretion rates are scaled by
$\Mstar$, and the optically thin disc masses assume
a common effective orbital distance rather than a
common average temperature for the emitting dust grains.
That is, the characteristic (mass-weighted) temperature
of the emitting grains, when averaged over the radial structure
of the disc, depends on $\Lstar$; in contrast, the
Andrews \& Williams (2005) disc masses used in Najita et al.\ (2007)
assumed a characteristic grain temperature of 20\,K
for all sources
(for details see section 3.2.2 of Andrews et al.\ 2013).

As shown in Figure 2, with either set of disc mass estimates,
the normal (single) T Tauri stars in
Oph overlap the region of the diagram occupied by the
normal (single) T Tauri stars in Tau.
Among the sources with measured stellar accretion rates and
disc masses,
there is a trend of $\Mdot$ increasing with $\Mdisk,$
albeit with significant dispersion.
A similar result was noted previously for the
Tau sample alone (Najita et al.\ 2007). A trend of $\Mdot$
increasing with $\Mdisk$ is not surprising;
in $\alpha$-viscosity discs (e.g., Shakura \& Sunyaev 1973)
the accretion rate is proportional to the disc column density
(or mass).

Time variability and measurement uncertainty in the stellar accretion
rate likely account for some of the observed scatter about
the mean trend.
Stellar accretion rates of individual objects are known to 
vary by a factor of $\sim 2$
(e.g., Hartigan et al.\ 1991).
Gullbring et al.\ (1998) reported typical measurement
uncertainties of 0.5 dex in their stellar accretion rates;
Natta et al.\ (2006) did not report their uncertainties.

The assumptions made in deriving disc masses are also
a likely source of dispersion. Disks may be partially
optically thick or have different temperature distributions
than assumed in the prescription of Andrews et al.\ (2013).
In addition, $\Mdisk$ measurements are insensitive to solids
that have grown into large grains ($\ga$ cm).
Dispersion in the solid size distribution would mean dispersion
in the extent to which the millimeter continuum traces the gas mass.
Unrecognized binarity (and its impact on either or both of
$\Mdot$ and $\Mdisk$) is yet another potential source of dispersion,
although more so for Oph than for Tau. The binarity of the latter population
is now fairly well known (e.g., Kraus et al.\ 2011).

Several additional Oph sources have low $\Mdot$ for their $\Mdisk$
but lack obvious MIR dips in their SED (Fig.~2, red open circles).
DoAr 25, GY284, and WL14
have upper limits below $\log\Mdot=-9.5,$ and
LFAM3 has the lowest Oph $\log\Mdot$ measurement in the plot.
WL14 and LFAM3 have high estimated extinction
values ($A_v>15$), with their stellar accretion rates possibly underestimated
as a result.
We discuss this issue and the possible nature of these
``outlier'' sources in the next section.

We also find that the known Oph transition discs occupy roughly
the same region of the $\Mdot$--$\Mdisk$ plane as the Tau
transition discs.
That is, the Oph transition discs have lower accretion rates
for their disc masses than do normal T Tauri stars in
Tau and Oph.
In Figure 2a,
the median $\log\Mdot$ for the normal T Tauri stars 
(outliers excluded) and
transition discs is -7.6 and -8.4, respectively; 
the median $\log\Mdisk$ is -2.3 and -1.7, respectively. 
If the outliers were included in the normal T Tauri 
population, the median $\log\Mdot$ would decrease slightly 
to $-7.8$; the median $\log \Mdisk$ would be unchanged.

With the addition of the Oph sources, we can describe more
quantitatively the trend between $\log\Mdisk$ and $\log\Mdot$
for the normal T Tauri population (outliers excluded) in Figure 2a.
The Kendall $\tau$ for the individual Tau and Oph and
the combined (Tau+Oph)
normal T Tauri samples is 0.34, 0.46, and 0.38, respectively,
with corresponding 2-sided $p$-values of
0.96\%, 2.8\%, and 0.03\%.
The reduced $p$-value of the combined sample illustrates the
value of the Oph sources in improving the statistical
significance of the trend.
Including the outliers in the normal T Tauri population 
reduces the Kendall $\tau$ to 0.23 ($p$-value of 2.3\%), 
consistent with the visual impression of these sources as 
outliers.  

We estimated the slope of the trend using
Brandon Kelly's {\it linmix\_err idl} routine
(Kelly 2007), which takes
a Bayesian approach to linear regression given errors in both
quantities.
Assuming errors of 0.2 dex in $\log\Mdiskthin$ and 0.3 dex in
$\log\Mdotnorm$, we find
\begin{equation}
\log\Mdotnorm = -5.15(\pm0.75) + 1.08(\pm0.3)\log\Mdiskthin
\end{equation}
with a dispersion about this trend of
$\sigma(\log\Mdotnorm)=0.5.$
If the Tau and Oph normal T Tauri star populations are examined separately,
they have similar slopes ($\sim 1$) and dispersion ($\sim 0.5$ dex).
The $\Mdotnorm$ normalization constant for the Oph normal T Tauri
population is larger by 0.75 dex
than that of the Tau population, consistent with the visual impression,
although the uncertainty in the values of the constants is large.

The small sample size of the transition discs makes it difficult
to compare in detail the distributions of the normal T Tauri stars and
transition disc populations in the
$\log\Mdisk$-$\log\Mdot$ plane.
For example, if we were to fit a linear trend to the transition
disc sample for comparison with the fit for the normal T Tauri stars
(eq.~2), the uncertainty on the slope and constant
would be too large for a meaningful comparison.
However, we can explore whether the distributions of the two populations
in the $\Mdot$--$\Mdisk$ plane are grossly similar
by dividing the plane
into two regions and seeing whether there is a statistical difference
in the distribution of the two populations across the two regions.
Using the distribution of normal T Tauri stars as a guide, since it is
the better characterized population, we divide the plane
at a line located 1-$\sigma$ below
the mean $\log\Mdiskthin$-$\log\Mdotnorm$
relation for the Tau+Oph normal T Tauri stars (eq.~2),
i.e., at
\begin{equation}
\log\Mdotnorm = -5.65 + 1.08\log\Mdiskthin
\end{equation}
(Fig.~2a, dashed line).
As expected, many more of the normal T Tauri stars in the combined sample lie
above the line (32 sources) than below the line (12 sources).
In contrast, there are no transition discs above the line and
8 below.
For the purpose of apportioning the sample,
upper limits in $\log\Mdot$ are treated as detections,
although the results are insensitive to how they are treated.

We can use a contingency table analysis (e.g., Bussmann et al.\ 2011)
to see whether these
values are consistent with the null hypothesis that the populations
are similarly distributed above and below the line.
Fisher's exact test, which is a useful approach when the sample sizes are
small, gives a probability of 0 that the T Tauri stars
and transition discs are similarly distributed.
The result is insensitive to where the boundary (eq.~3) is drawn.
If the boundary were a factor of 2, 2.5, or 3 times higher in
$\log\Mdot$, the probability would still be only 0.002, 0.006, or 0.014
that the T Tauri stars and transition discs are similarly
distributed. The boundary in the last case is coincident with the
locus of the mean trend (Fig.~2a, dotted line).
If the boundary were a factor of 2 or 3 lower in $\log\Mdot$,
the probability remains low at 0.001 or 0.002 respectively.
Including the outliers in the normal T Tauri population 
raises the above probabilities slightly (by a factor of $\sim 2$)  
but does not change our finding that the 
transition discs and normal T Tauri stars are not similarly 
distributed in the $\Mdot$--$\Mdisk$ plane. 
As a contrasting example, we can also compare the distributions
of the Tau and Oph normal T Tauri stars across the boundary
(eq.~3).
The probability that the two populations are similarly
distributed is 21\%.

\section{Discussion} 
\label{sec: discuss}

As described in the previous section,
when combined with the Tau sample, the Oph sample enhances the
statistical signficance of the previously reported trend of
$\Mdot$ vs.\ $\Mdisk$ among single, normal T Tauri stars.
The Oph transition discs, like the Tau transition discs,
are found to
favor the lower right region of the $\Mdot$-$\Mdisk$ plane,
i.e., transition discs tend to have lower $\Mdot$ for a given
$\Mdisk$ and a higher $\Mdisk$ (for a given $\Mdot$) than
normal T Tauri stars.

Planetesimal formation does not provide a ready explanation for
this result.
Naively, we would expect that systems forming planetesimals would
have accretion rates similar to those of normal T Tauri stars,
because planetesimal formation alters the size distribution of
the disc solids, but it is not known to alter
the properties of the disc gas and stellar accretion rate.
While this result was obtained previously for the Tau
transition disc sample,
it also appears to apply at the younger average age of Oph, when
planetesimal formation, as the first step in planet formation in the core accretion scenario,
would be even more likely to dominate over the other proposed explanations for
transition discs (giant planet formation, photoevaporation).
The lack of transition discs meeting our expectations for where planetesimal-forming
systems would lie in the $\Mdot$-$\Mdisk$ plane,
for both the Tau and Oph samples,
argues against planetesimal formation as a significant pathway to form
transition discs at the ages of nearby star forming regions ($1-3$\,Myr).

This interpretation is consistent with the results of recent
theoretical studies that find it difficult to replicate the
properties of transition discs through grain evolution
alone.  Early models of grain evolution that considered sticking
without fragmentation were able to produce
MIR SED dips like those of transition discs
(Dullemond \& Dominik 2005; Tanaka et al.\ 2005).
However, recent, more realistic grain evolution models that include
coagulation, fragmentation, radial drift, gas drag, and
turbulent mixing (Birnstiel et al.\ 2012) find that while discs
can develop the MIR dips that characterize transition disc SEDs,
enough non-fragmenting centimeter-sized particles remain behind
that these systems would not show the submillimeter cavities
that also characterize transition discs.
A similar difficulty is found even under the simplified conditions
considered in earlier models (perfect sticking in the absence
of radial draft; e.g., Dullemond \& Dominik 2005).

The measurable $\Mdot$ and higher average $\Mdisk$ of
transition discs compared to normal T Tauri stars
also suggests that photoevaporation is not the dominant
pathway to a transition disc at Taurus-Ophiuchus age.
Photoevaporation is more likely to create an inner cavity in the
disc in systems with low $\Mdot$ and low $\Mdisk$, if all
other factors are equal (e.g., Alexander \& Armitage 2007).

The low $\Mdot$ of transition discs for their disc masses
relative to normal T Tauri stars is most consistent with the formation
of objects massive enough to alter the disc accretion flow, i.e.,
giant planet formation.
Interestingly, this is the case for both systems with transition disc
SEDs and those with submillimeter cavities.
Sources such as DoAr44, a transition disc
with a prominent MIR SED dip, are hypothesized to have formed
giant planetary companion(s) (e.g., Espaillat et al.\ 2010).

Why do we find low accretion rates for transition discs both
with and without MIR dips in their SEDs?
Perhaps the conditions needed to reduce the stellar accretion rate
are less stringent than those needed to render the inner disc optically
thin in the MIR.
Removing small dust grains from the inner disc enough to render
it optically thin while also allowing continued accretion onto
the star is a recognized challenge (Rice et al.\ 2006; Zhu et al.\ 2012).
Whether a system with a submillimeter cavity also develops an
optically thin inner disc in the MIR (i.e., a transition disc
SED) may therefore require the formation of higher mass companions and/or
a larger number of companions
(Zhu et al.\ 2011; Dodson-Robinson \& Salyk 2011).
For example, a more massive companion or more of them may induce a
larger pressure bump at the inner edge of the outer disc,
enhancing the efficiency of dust filtering (Rice et al.\ 2006).
Radiation pressure from a high mass accreting planet can also play a
role in creating a transition disc SED (Owen 2014).
In comparison, the formation of a single,
low mass giant planet
 may be sufficient to create a submillimeter cavity and
to impact the disc accretion rate (e.g., Lubow et al.\ 1999).

In \S 3, we noted a few outliers among the Oph sources in the
$\Mdot$-$\Mdisk$ plot,
i.e., sources with normal T Tauri SEDs but low accretion rates for their
disc mass. For some sources, their high $A_V$ may have led to
an underestimate of their accretion rates (WL14, LFAM3).
In sources with very high reddening,
such as highly-inclined discs or protostars, the scattered light component can
dominate the observed light into the near-infrared.
Because the colors then appear bluer than would be produced by
the extinction along the line of sight to the central source,
the typical methods of determining
the extinction using observed near-infrared colors and assuming
an intrinsic color from a known
spectral type or the classical T Tauri star locus will underestimate
the extinction.
Subsequent
dereddening of the observed line flux will lead to an underestimate of the true
line luminosity, and by extension, $\Mdot$.
Correcting for this effect for a given source
depends on the scattering fraction, which is difficult to determine.

Another possibility is that some or all of the low accretion rate sources
may be unrecognized transition discs in which giant planet formation
either {\it (i)}
has occurred at disc radii too small to probe
with SEDs or submillimeter imaging
or
{\it (ii)}
is insufficiently advanced (e.g., the planet is too low in mass or
too few in number) to render portions of the disc optically thin.
This scenario may be particularly relevant to the
two outliers that have only modest
extinction (DoAr25 at $A_v=3.4$ and GY284 at $A_v=6$).

Andrews et al.\ (2008) previously noted that DoAr25
has a low stellar accretion rate for its disc mass.
Although DoAr25 has one of the highest measured
disc masses ($0.14 \Msun$),
stellar accretion estimates range from
$< 2\times 10^{-10}\Msunperyr$ (Natta et al.\ 2006)
to $3\times 10^{-9}\Msunperyr$ (Greene \& Lada 1996; Luhman \& Rieke 1999).
Transition discs like LkCa15 and UX TauA,
which also have low accretion rates for their disc masses,
have such large gaps ($\ga 25$\,AU) that multiple giant planets,
spread out over a range of radii,
have been invoked to explain the large optically thin regions
(Zhu et al.\ 2011).
In contrast, T Tauri discs that have formed a single giant planet
are expected to clear much smaller gaps,
particularly if the planet is located at small disc radii,
e.g., at a Jupiter-like distance of 5\,AU.
A high mass giant planet ($\sim 5M_J$) located at such a distance
is expected to greatly reduce the stellar accretion rate
(Lubow et al.\ 1999).
Given its large disc mass ($\sim 0.1\Msun$),
it seems possible that DoAr25 may have formed such a giant planet.
Interestingly, the accretion rate estimates in the literature
for DoAr25 differ by more than a factor of 10. Extreme variability
(by such a large factor)
in the stellar accretion rate is unusual (e.g., Nguyen et al.\ 2009)
and may be consistent with the presence of a massive close companion
(e.g., Basri et al.\ 1997).

Other sources near Oph may share these properties. For example, 
the MIR colors of RXJ1633.9-2242 (source \#32 in Cieza et al.\ 2010) 
are consistent with the presence of a disk cavity $\sim 8$\,AU in 
radius (Orellana et al.\ 2012). 
The source has a fairly large disk mass 
($\Mdisk \simeq$ 0.01--0.02\,$\Msun$) and 
a low estimated accretion rate based on its H$\alpha$ width 
($\Mdot = 10^{-10}\Msunperyr$). The low accretion rate could be 
confirmed with a more accurate accretion rate diagnostic
(e.g., IR hydrogen line luminosities or $U$-band excess). 

Cavities 5\,AU or 8\,AU in radius would have been difficult
to detect in the submillimeter studies of nearby young stars
reported to date.
The smallest dust cavity reported in the SMA study of
Andrews et al.\ (2011) has a radius of 15\,AU (WSB60).
Higher angular resolution observations (e.g., with ALMA) of
DoAr25 and the other low accretion rate systems
would explore whether these systems are like the known Oph
transition discs in having inner cavities.

Thus we find that the combination of stellar accretion rates
and disc masses can help to distinguish among proposed scenarios
for the origin of transition discs.
Toward this end, we have attempted to control for the known
dependencies of 
$\Mdot$ and $\Mdisk$ on age (e.g., Hartmann et al.\ 1998), 
$\Mdot$ on stellar mass (e.g., Muzerolle et al.\ 2003), 
and $\Mdisk$ on the presence of stellar or sub-stellar
companions 
(e.g., Artymowicz \& Lubow 1996; Lubow et al.\ 1999) by
{\it (i)} comparing the properties of transition discs and normal
T Tauri stars in the same star-forming region, 
{\it (ii)} correcting for the known dependence of $\Mdot$ on $\Mstar$, and
{\it (iii)} excluding known binaries, respectively.
With this approach, we find that including disc masses in
the demographic analysis helps to accentuate decrements
in the stellar accretion rates of transition disc subpopulations
relative to normal T Tauri stars.

Our results complement previous studies that have taken different
approaches in examining the accretion rates of transition discs
in other star forming regions. Some of the earlier studies focused
on sources older than those studied here. For example, in the
Tr 37 and NGC\,7160 clusters in Cep OB2 ($\sim 4$\, Myr and
$\sim 12$\,Myr respectively), half the transition discs
exhibit no measurable accretion,
and
the accretion rates of the other half are similar to those of
classical T Tauri stars
in the region ($2-3\times 10^{-9}\Msunperyr$; Sicilia-Aguilar et
al.\ 2010). These results may differ from our result
because the dominant physical mechanism that produces transition
disc SEDs may differ between ages of 1\,Myr and 10\,Myr; e.g.,
photoevaporation may dominate at late times.

Several studies have compared the stellar accretion rates
(but not disc masses) of transition discs and T Tauri stars
in star forming regions similar in age to those
studied here ($\la 3$\,Myr).
Similar in spirit to our results, some studies have
reported evidence that transition discs have lower accretion
rates (Espaillat et al.\ 2012) or lower accretion luminosities
(Salyk et al.\ 2013) than normal T Tauri stars.
Kim et al.\ (2013) found that
transition discs in Orion A have accretion rates
a factor of $\sim 10$ lower, on average, than Taurus
T Tauri stars of the same stellar mass.
In contrast, Fang et al.\ (2013) found that accreting
transition discs in L1641 have
a median accretion rate (from H$\alpha$) similar to that of
the T Tauri stars in the region.
However, their transition disc accretion rates determined 
from H$\beta$
luminosity do have a lower median value (by a factor $\sim 3$)
than T Tauri stars of the same stellar mass.
H$\beta$ luminosity may more reliably trace accretion rate than
H$\alpha,$ because H$\alpha$ is more optically thick and it can
be reduced in strength by wind absorption components
(e.g., Muzerolle et al.\ 1998).

The choice of comparison samples may also play a role in the
the interpretation of accretion rates.
Manara et al.\ (2014) compared the accretion rates
of transition discs in a variety of star forming regions
(Taurus, Ophiuchus, Chamaeleon, Serpens) with the accretion rates
of T Tauri stars in Lupus I and II.
The transition discs appear biased toward lower $\Mdot$,
with a median $\Mdot$ $\sim 3$ times lower than the average rate
for the T Tauri stars for the same $\Mstar$.
The use of Lupus T Tauri stars as the comparison sample
may underestimate the $\Mdot$ decrement, because
stellar accretion rate declines with age (Hartmann et al.\ 1998),
and the Lupus I and II regions ($\sim 3$\,Myr; Alcal\'a et al.\ 2014)
may be older than most of the transition discs in the Manara et al.\ study
(1--2\,Myr for Tau, Oph, Cha; e.g., Furlan et al.\ 2009).
The average $\Mdot$ for Lupus T Tauri stars
(0.4--0.8\,$\Msun$) is $1.3\times 10^{-9} \Msunperyr$ (Alcal\'a et al.\
2014), whereas the mean $\Mdot$ for Taurus T Tauri stars is $\sim
10^{-8}\,\Msunperyr$ (Hartmann et al.\ 1998).
Thus, other studies in the literature appear to find results
consistent with those reported here when we consider the
accretion rate tracers and comparison populations used.

\section{Summary and Conclusions} 

We find that the combination of stellar accretion rates
and disc masses can aid in distinguishing among proposed
explanations for the unusual SEDs of the population of 
young stellar objects known as transition discs.
Determining how stellar accretion rate
depends on a parameter such as SED shape can be a challenge
because accretion rate is known to depend on other parameters
such as age (e.g., Hartmann et al.\ 1998),
stellar mass (e.g., Muzerolle et al. 2003),
disc mass (Najita et al.\ 2007),
and potentially the presence of stellar or sub-stellar
companions (e.g., Artymowicz \& Lubow 1996; Lubow et al.\ 1999).
In our study, we have attempted to control for these
known dependencies by
{\it (i)} comparing the properties of transition discs and normal
T Tauri stars in the same star-forming region,
{\it (ii)} correcting for the known dependence of $\Mdot$ on $\Mstar$,
{\it (iii)} using available $\Mdisk$ measurements, and
{\it (iv)} excluding known binaries, respectively.

Using this approach, we compared the distributions
in the $\Mdot$-$\Mdisk$ plane
of transition discs and single, normal
T Tauri stars in Taurus and Ophiuchus.
The transition discs we studied have prominent MIR dips in
their {\it Spitzer} SEDs or inner cavities in submillimeter
continuum images.
The available data suggest some interesting trends:

1.  When combined with the Tau sample, the Oph sample enhances
the statistical signficance of the previously reported trend of
$\Mdot$ vs.\ $\Mdisk$ among single, normal T Tauri stars.

2.  The Tau and Oph transition discs we studied
favor the lower right region of the $\Mdot$-$\Mdisk$ plane,
i.e., they tend to have lower $\Mdot$ for a given
$\Mdisk$ and a higher $\Mdisk$ (for a given $\Mdot$) than
normal T Tauri stars.
This result is most consistent with the formation of objects
massive enough to alter the accretion flow, i.e., single or multiple
giant planets.

3. Several T Tauri stars that are not known transition disc systems also
have low accretion rates for their disc masses.  Such sources that
also have
high extinctions may have underestimated stellar accretion rates
(LFAM3, WL14).  The others may have formed (high mass) giant planets
that are massive enough to greatly reduce the stellar accretion rate,
but the planets are
located at orbital radii too small to have an observable impact
on the submillimeter continuum image or a distinctive impact on the SED   
(DoAr25, GY284).  More
detailed studies of the extinction toward these sources, their
stellar accretion rates, and disc structure would help to distinguish
between these possibilities.

\section*{Acknowledgments}

We are grateful to Melissa McClure for providing the stellar luminosities
for the Oph sources studied here.
JN is grateful to the Institute for Theory and Computation at the
Harvard-Smithsonian Center for Astrophysics for supporting a
sabbatical visit that enabled this study.

\begin{table*}
\begin{minipage}{7.truein}
 \caption{Ophiuchus Stellar Accretion Rates and Disk Masses}
\begin{tabular}{@{}llccccrrclrcl}
\hline
  Source        & SpT
                        & Mult
                        & $\Lstar$
                        & $log\Rstar$
                        & $log\Mstar$
                        & $log\Mdot$
                        & $log\Mdisk$
                        & $log M_{\rm d,fit}$
                        & Ref 
                        & $A_V$
                        & Type
                        & Other Name \\
(1)     & (2)
        & (3)
        & (4)
        & (5)
        & (6)
        & (7)
        & (8)
        & (9)
        & (10)
        & (11)
        & (12)
        & (13) \\
 \hline
   IRS2 & K3.5 & B &  1.66 &  0.29 &  0.16 & $<$ -7.29 & $<$ -3.58 &       &      & 16.9 &  N &              YLW19\\
    SR4 & K4.5 & S &  1.31 &  0.28 &  0.08 &     -7.28 &     -2.37 &       &      &  2.2 &  N &        AS206/IRS12\\
  GSS26 &   K7 & S &  8.14 &  0.76 & -0.09 & $<$ -7.73 &     -2.13 &       &      & 21.9 &  N &                  -\\
   EL18 &   M1 & B &  2.14 &  0.55 & -0.33 & $<$ -8.01 & $<$ -3.14 &       &      &  9.9 &  N &              GSS29\\
 DoAr24 &   K5 & S &  1.23 &  0.29 &  0.02 &     -8.88 & $<$ -2.59 &       &      &  3.5 &  N &       GSS28/HBC638\\
   EL20 &   M0 & S &  1.97 &  0.50 & -0.27 &     -6.82 &     -2.15 & -1.54 & A09  & 14.2 &  N &              VSSG1\\
   GY11 &   M6 & S & 0.002 & -0.78 & -1.13 & $<$-12.68 & $<$ -2.48 &       &      &  4.2 &  N &                  -\\
 DoAr25 &   K5 & S &  1.52 &  0.33 &  0.04 & $<$ -9.55 &     -1.71 & -0.87 & A09  &  3.4 &  O &         WSB29/GY17\\
  LFAM3 &   M0 & S &  0.31 &  0.09 & -0.24 &     -8.90 &     -1.83 &       &      & 16.1 &  O &               GY21\\
   EL24 &   K5 & S &  5.10 &  0.60 &  0.04 &     -6.43 &     -1.65 & -0.93 & A09  &  8.7 &  N &              WSB31\\
 VSSG27 &   K6 & S &  1.83 &  0.40 & -0.04 &     -7.02 & $<$ -2.82 &       &      & 24.0 &  N &               GY51\\
  WSB37 &   M5 & B &  0.21 &  0.19 & -0.69 & $<$ -9.58 & $<$ -2.42 &       &      &  2.5 &  N &               GY93\\
  IRS17 &   K3 & S &  1.14 &  0.20 &  0.17 & $<$ -9.33 & $<$ -2.58 &       &      & 20.3 &  N &   WL7/GY98/WLY2-17\\
   EL26 &   M0 & S &  1.71 &  0.47 & -0.28 &     -8.04 &     -2.93 &       &      &  7.5 &  N &      GSS37AB/GY110\\
   EL27 &   M0 & S &  1.28 &  0.40 & -0.28 &     -7.10 &     -1.57 & -0.84 & A09  & 14.8 &  N &        GSS39/GY116\\
  WSB38 & G3.5 & B &  5.46 &  0.36 &  0.07 &     -8.64 & $<$ -2.95 &       &      &  7.0 &  N &     VSS27AB/ROXs16\\
   WL18 & K6.5 & S &  0.27 &  0.00 & -0.10 &     -7.84 &     -2.45 &       &      & 10.5 &  N &              GY129\\
   WL14 &   M4 & S &  0.14 &  0.06 & -0.62 & $<$-10.20 &     -2.28 &       &      & 15.8 &  O &              GY172\\
   WL16 &   A1 & S &250.   &  0.79 &  0.65 &     -5.26 & $<$ -3.93 &       &      & 31.0 &  N &         GY182/YLW5\\
  IRS29 &   M4 & B &  0.63 &  0.39 & -0.58 & $<$ -9.90 &     -3.18 &       &      & 17.8 &  N &        WL1AB/GY192\\
   WL10 &   K7 & S &  0.96 &  0.29 & -0.12 &     -8.63 & $<$ -3.03 &       &      & 11.5 &  N &              GY211\\
   SR21 &   F4 & S & 16.54 &  0.49 &  0.16 & $<$ -8.51 &     -2.40 & -2.22 &  A11 &  6.9 & HT &        EL30/VSSG23\\
   WL11 &   M0 & S &  0.10 & -0.15 & -0.27 &    -10.33 & $<$ -2.71 &       &      & 11.5 &  N &              GY229\\
  IRS34 &   M0 & B &  1.07 &  0.36 & -0.27 & $<$ -8.41 & $<$ -2.74 &       &      & 24.7 &  N &              GY239\\
  IRS39 & M1.5 & S &  1.05 &  0.41 & -0.38 & $<$ -8.97 &     -2.62 &       &      & 18.4 &  N &          WL4/GY247\\
  IRS42 &   K7 & S &  7.29 &  0.73 & -0.10 & $<$ -7.57 & $<$ -2.51 &       &      & 25.0 &  N &              GY252\\
 YLW16C &   M1 & S &  1.39 &  0.45 & -0.34 &     -7.49 &     -2.30 &       &      & 22.5 &  N &              GY262\\
   EL31 &   M4 & B &  0.87 &  0.46 & -0.58 & $<$ -8.59 & $<$ -3.02 &       &      & 11.7 &  N &  WL13/VSSG25/GY267\\
   EL32 & K6.5 & S &  1.75 &  0.41 & -0.07 & $<$ -8.65 & $<$ -2.41 &       &      & 21.1 &  N &        IRS45/GY273\\
   EL33 &   M3 & B &  4.88 &  0.80 & -0.47 & $<$ -7.70 & $<$ -2.94 &       &      & 24.1 &  N &        IRS47/GY279\\
  GY284 & M3.2 & S &  0.12 &  0.00 & -0.56 & $<$-10.31 &     -1.62 &       &      &  6.0 &  O &                  -\\
 ROXs25 &   K7 & S &  3.07 &  0.55 & -0.11 &     -7.20 &     -2.71 &       &      & 11.6 &  N &              GY292\\
  IRS49 & K5.5 & S &  2.16 &  0.43 &  0.22 &     -8.30 &     -2.70 &       &      & 10.5 &  N &              GY308\\
  WSB52 &   K5 & S &  0.60 &  0.27 & -0.34 &     -8.26 &     -2.15 & -2.15 &  A10 &  5.0 &  N &              GY314\\
    SR9 &   K5 & B &  2.10 &  0.40 &  0.04 &     -8.43 & $<$ -3.12 &       &      &  1.6 &  N &  AS207/GY319/IRS52\\
   SR10 &   M2 & S &  0.33 &  0.18 & -0.43 &     -8.30 & $<$ -2.49 &       &      &  1.3 &  N &       GY400/HBC265\\
  WSB60 & M4.5 & S &  0.20 &  0.16 & -0.65 &     -8.59 &     -1.74 & -1.55 &  A11 &  3.9 &  H &YLW58/B162816-243657\\
   SR20 &   G7 & B &  8.71 &  0.49 &  0.07 & $<$ -8.41 & $<$ -3.30 &       &      &  6.0 &  N &             HBC643\\
   SR13 &   M4 & B &  0.40 &  0.29 & -0.59 &     -8.46 &     -2.09 & -1.92 &  A10 &  0.0 &  N &             HBC266\\
 AS205A &   K5 & S &  4.00 &  0.54 &  0.04 &     -6.47 &     -1.71 & -1.54 &  A10 &  2.9 &  N &                  -\\
  AS209 &   K5 & S &  1.50 &  0.33 &  0.04 &     -7.29 &     -1.76 & -1.60 &  A10 &  0.9 &  N &                  -\\
 DoAr44 &   K3 & S &  1.90 &  0.31 &  0.20 &     -8.04 &     -2.11 & -2.15 &  A11 &  3.3 & HT &             ROXs44\\
  SR24S &   K2 & S &  4.40 &  0.46 &  0.30 &     -7.15 &     -1.88 & -1.35 &  A11 &  7.0 &  H &                  -\\
 WaOph6 &   K6 & S &  2.90 &  0.50 & -0.04 &     -6.64 &     -2.02 & -1.11 &  A10 &  3.6 &  N &                  -\\
\hline 
\end{tabular}
\medskip
Column 2: Spectral types from McClure et al.\ (2010),
Andrews \& Williams (2007a, 2007b), Andrews et al.\ (2009, 2010).
Column 3: Sources with a companion within $1\arcsec$ are classified as binaries
(Reipurth \& Zinnecker 1993, Barsony et al.\ 2005; Ratzka et al.\ 2005).
Columns 4 and 12: Stellar luminosity (in $\Lsun$) and $A_v$ from 
McClure et al.\ (2010) and M. McClure (2014, private communication) except for 
WL16 (Ressler \& Barsony 2003), 
WL14 (Luhman \& Rieke 1999), 
GY11 (Cushing et al.\ 2000), 
EL24, WSB52, SR13, AS205A, AS209, SR24S, WaOph6 (Andrews et al.\ 2010).
Columns 5 and 6: Derived stellar radius (in $\Rsun$) and stellar mass (in $\Msun$).
Column 7: Stellar accretion rate derived from Natta et al.\ (2006) measurements using the stellar parameters from McClure et al.\ (2010) (see text for details).
Column 8: Disk mass (in $\Msun$) estimated from optically thin, isothermal approximation (see \S2 for details).
Column 9: Alternate disc mass estimate from a fit to the SED and resolved millimeter visibilities.
Column 10: Reference for $\Mdiskfit$:
A09 = Andrews et al.\ (2009);
A10 = Andrews et al.\ (2010);
A11 = Andrews et al.\ (2011).
Column 12: Source type, where 
N= normal T Tauri star;
T= transition disc SED;
H= submillimeter cavity;
O= outlier.
\end{minipage}
\end{table*}

\begin{table*}
\begin{minipage}{4.5truein}
 \caption{Taurus Stellar Accretion Rates and Disk Masses}
 \begin{tabular}{@{}lccccccl}
\hline 
Source        & SpT
                        & $\log\Mstar$
                        & $\log\Mdot$
                        & $\log\Mdisk$
                        & $\log M_{\rm d,fit}$
                        & Ref $\Mdiskfit$
                        & Type \\
(1)     & (2)
        & (3)
        & (4)
        & (5)
        & (6)
        & (7)
        & (8) \\
\hline
   AATau &   K7   &  -0.10 &    -8.48 &    -2.04 & -1.52 & A07 & N \\
   BPTau &   K7   &  -0.10 &    -7.54 &    -2.25 &       &     & N \\
   CITau &   K7   &  -0.10 &    -7.19 &    -1.84 & -1.40 & A07 & N \\
   CWTau &   K3   &   0.20 &    -7.61 &    -2.27 &       &     & N \\
   CXTau &   M2.5 &  -0.44 &    -8.97 &    -2.77 &       &     & N \\
   CYTau &   M1.5 &  -0.36 &    -8.12 &    -1.85 &       &     & N \\
   DETau &   M1   &  -0.41 &    -7.59 &    -2.39 &       &     & N \\
   DGTau &   K7   &  -0.04 &    -6.30 &    -1.43 &       &     & N \\
  DHTauA &   M1   &  -0.32 &    -8.30 &    -2.54 & -2.52 & A07 & N \\
   DKTau &   M0   &  -0.15 &    -7.42 &    -2.69 &       &     & N \\
   DLTau &   K7   &  -0.09 &    -6.79 &    -1.61 & -1.00 & A07 & N \\
   DMTau &   M1   &  -0.33 &    -7.95 &    -1.72 & -1.40 & A11 & T \\
   DNTau &   M0   &  -0.22 &    -8.46 &    -1.93 & -1.22 & A07 & N \\
   DOTau &   M0   &  -0.33 &    -6.85 &    -1.88 &       &     & N \\
   DRTau &   K7   &   0.08 &    -6.50 &    -1.89 & -2.00 & A07 & N \\
   DSTau &   K5   &   0.02 &    -7.89 &    -2.62 &       &     & N \\
   FMTau &   M0   &  -0.22 &    -8.45 &    -2.63 &       &     & N \\
   FYTau &   K7   &   0.07 &    -7.41 &    -2.80 &       &     & N \\
   FZTau &   M0   &  -0.35 &    -7.32 &    -2.77 &       &     & N \\
   GITau &   K7   &  -0.10 &    -7.69 &    -2.81 &       &     & N \\
   GKTau &   K7   &  -0.11 &    -8.19 &    -3.44 &       &     & N \\
   GMAur &   K3   &   0.13 &    -8.02 &    -1.66 & -1.15 & A11 & T \\
   GOTau &   M0   &  -0.20 &    -7.93 &    -1.98 & -0.74 & A07 & N \\
   HKTau &   M0.5 &  -0.27 &    -7.27 &    -2.27 &       &     & N \\
   HNTau &   K5   &  -0.04 &    -8.41 &    -2.64 &       &     & N \\
   HOTau &   M0.5 &  -0.25 &    -8.86 &    -2.34 &       &     & N \\
   IPTau &   M0   &  -0.22 &    -9.10 &    -2.83 &       &     & N \\
   IQTau &   M0.5 &  -0.27 &    -7.55 &    -2.06 &       &     & N \\
  LkCa15 &   K5   &   0.02 &    -8.87 &    -1.74 & -1.26 & A11 & T \\
  RWAurA &   K3   &   0.18 &    -7.12 &    -2.52 &       &     & N \\
   RYTau &   K1   &   0.44 &    -7.11 &    -1.92 & -1.70 & A07 & N \\
  UXTauA &   K2   &   0.20 &    -9.00 &    -2.22 & -2.15 & A11 & T \\
 V836Tau &   K7   &  -0.08 &    -8.98 &    -2.33 &       &     & N \\
\hline
\end{tabular}
\medskip
Column 3: Stellar mass (in $\Msun$).
Column 4: Stellar accretion rate (in $\Msunperyr$).
Column 5, 6: Disk mass (in $\Msun$).
Column 7: A07 = Andrews \& Williams (2007b);
A11 = Andrews et al.\ (2011).
Column 8: Source type, where
N= normal T Tauri star;
T= transition disc.
\end{minipage}
\end{table*}

\bsp

\label{lastpage}

\end{document}